\documentclass[11pt]{article}
\usepackage{moriond,epsfig}

\def\be{\begin{equation}}
\def\ee{\end{equation}}
\def\bea{\begin{eqnarray}}
\def\eea{\end{eqnarray}}

\newcommand{\gsim}{\mbox{${~\raise.25em\hbox{$>$}\kern-.70em
\lower.25em\hbox{$\sim$}~}$}}
\newcommand{\lsim}{\mbox{${~\raise.25em\hbox{$<$}\kern-.70em
\lower.25em\hbox{$\sim$}~}$}}

\begin{document}
\vspace*{4cm}
\title{RECENT ISSUES IN LEPTOGENESIS}

\author{ ENRICO NARDI }

\address{
INFN, Laboratori Nazionali di Frascati, C.P. 13,
      100044 Frascati, Italy,  and \\
    Instituto de F\'{i}sica, Universidad de Antioquia,
    A.A.1226, Medell\'{i}n, Colombia
}

\maketitle\abstracts{ Baryogenesis via leptogenesis provides an appealing
  mechanism to explain the observed baryon asymmetry of the Universe.  Recent
  refinements in the understanding of the dynamics of leptogenesis include
  detailed studies of the effects of lepton flavors and of the role possibly
  played by the lepton asymmetries generated in the decays of the heavier
  singlet neutrinos $N_{2,3}$. A review of these recent developments in the
  theory of leptogenesis is presented.  }

\section{Introduction}

The possibility that the Baryon Asymmetry of the Universe (BAU) could
originate from a lepton number asymmetry generated in the $CP$ violating
decays of the heavy seesaw Majorana neutrinos was put forth about twenty years
ago by Fukugita and Yanagida.~\cite{Fukugita:1986hr} Their proposal came
shortly after Kuzmin, Rubakov and Shaposhnikov pointed out that above the
electroweak phase transition $B+L$ is violated by fast electroweak anomalous
interactions.~\cite{Kuzmin:1985mm} This implies that any lepton asymmetry
generated in the unbroken phase would be unavoidably converted in part into a
baryon asymmetry. However, the discovery that at $T\gsim 100\,$GeV electroweak
interactions do not conserve baryon number, also suggested the exciting
possibility that baryogenesis could be a purely standard model (SM)
phenomenon, and opened the way to electroweak
baryogenesis.~\cite{Farrar:1993sp} Even if rather soon it became clear that
within the SM electroweak baryogenesis fails to reproduce the correct BAU by
many orders of magnitude,~\cite{Gavela:1993ts} within the minimal
supersymmetric standard model (MSSM) the chances of success were much better,
and this triggered an intense research activity in that direction. Indeed, in
the early 90's electroweak baryogenesis attracted more interest than
leptogenesis, but still a few remarkable papers appeared that put the first
basis for {\it quantitative} studies of leptogenesis. Here I will just mention
two important contributions that established the structure of the two main
ingredients of leptogenesis: the rates for several washout processes relevant
for the leptogenesis Boltzmann equations, that were presented by Luty in his
1992 paper,~\cite{Luty:1992un} and the correct expression for the $CP$
violating asymmetry in the decays of the lightest Majorana neutrino, first
given in the 1996 paper of Covi, Roulet and Vissani.~\cite{Covi:1996wh}

Around year 2000 a flourishing of detailed studies of leptogenesis begins,
with a corresponding burst in the number of papers dealing with this
subject.~\cite{Nardi:2007fs} This raise of interest in leptogenesis can be
traced back to two main reasons: firstly, the experimental confirmation (from
oscillation experiments) that neutrinos have nonvanishing masses strengthened
the case for the seesaw mechanism, that in turn implies the existence, at some
large energy scale, of lepton number violating ($\not \!  L$) interactions.
Secondly, the fact that the various analysis of supersymmetric electroweak
baryogenesis cornered this possibility in a quite restricted region of
parameter space, leaving for example for the Higgs mass just a 5 GeV window 
(115 - 120 GeV).~\cite{Balazs:2005tu}

The number of important papers and the list of people that contributed to the
development of leptogenesis studies and to understand the various implications
for the low energy neutrino parameters is too large to be recalled here.
However, let me mention the remarkable paper of Giudice {\it et
  al.}~\cite{Giudice:2003jh} that appeared at the end of 2003: in this paper a
whole set of thermal corrections for the relevant leptogenesis processes were
carefully computed, a couple of mistakes common to previous studies were
pointed out and corrected, and a detailed numerical analysis was presented
both for the SM and the MSSM cases.  Eventually, it was claimed that the
residual numerical uncertainties would probably not exceed the 10\%-20\%
level.  A couple of years later, Nir, Roulet, Racker and
myself~\cite{Nardi:2005hs} carried out a detailed study of additional effects
that were not accounted for in the analysis of ref..~\cite{Giudice:2003jh}
This included electroweak and QCD sphaleron effects, the effects of the
asymmetry in the Higgs number density, as well as the constraints on the
particles asymmetry-densities implied by the spectator reactions that are in
thermal equilibrium in different temperature ranges relevant for
leptogenesis.~\cite{Nardi:2005hs} Indeed, we found that the largest of theses
new effects would barely reach the level of a few tens of percent.

However, two important ingredients had been overlooked in practically all
previous studies, and had still to be accounted for.  These were the role of
the light lepton flavors, and the role of the heavier seesaw Majorana
neutrinos.  One remarkable exception was the 1999 paper by Barbieri {\it et
  al.}~\cite{Barbieri:1999ma} that, besides addressing as the main topic the
issue of flavor effects in leptogenesis, also pointed out that the lepton
number asymmetries generated in the decays of the heavier seesaw neutrinos can
contribute to the final value of the BAU.\footnote{Lepton flavor effects were
  also considered by Endoh, Morozumi and Xiong in their 2003
  paper,~\cite{Endoh:2003mz} in the context of the minimal seesaw model with
  just two right handed neutrinos.}  However, these important results did not
have much impact on subsequent analyses. The reason might be that these were
thought to be just order one effects on the final value of the lepton
asymmetry, with no other major consequences for leptogenesis. As I will
discuss in the following, the size of the effects could easily reach the one
order of magnitude level and, most importantly, they can spoil the
leptogenesis constraints on the neutrino low energy parameters, and in
particular the limit on the absolute scale of neutrino
masses.~\cite{buch02-03} This is important, since it was thought that this
limit was a firm prediction of leptogenesis with hierarchical seesaw
neutrinos, and that the discovery of a neutrino mass $m_\nu\gsim 0.2\,$eV
would have strongly disfavored leptogenesis, or hinted to different scenarios
(as e.g. resonant leptogenesis~\cite{pila0405}).

\section{The standard scenario}

Let us start by writing the first few terms of the leptogenesis Lagrangian,
neglecting for the moment the heavier neutrinos $N_{2,3}$ (except for their
virtual effects in the $CP$ violating asymmetries):
\begin{equation}
\label{lagrangian1}
{\cal L} =\frac{1}{2}\left[\bar N_1 (i\! \not\! \partial) N_1 - 
M_1 N_1 N_1\right] -(\lambda_{1}\,
\bar N_1\,  \ell_{1}{H} +{\rm h.c.}).
\end{equation}
Here $N_1$ is the lightest right-handed Majorana neutrino with mass
$M_1$, $H$ is the Higgs field, and $\ell_1$ is the lepton doublet to
which $N_1$ couples, that when expressed on a complete  orthogonal basis 
$\{\ell_i \}$  reads 
\begin{equation}
|\ell_1\rangle  = (\lambda  \lambda^\dagger )_{11}^{-1/2} 
\sum_i \lambda_{1i}\,|\ell_i\rangle.
\end{equation}
In practice it is always convenient to use the basis that diagonalizes the
charged lepton Yukawa couplings (the flavor basis) that also has well defined
$CP$ conjugation properties $CP(\{\ell_i \})=\{\bar \ell_i \}$ with
$i=e,\,\mu,\,\tau $.  Note that in the first and third term in
(\ref{lagrangian1}) a lepton number can be assigned to $N_1$, that is however
violated by two units by the mass term.  Then eq.~(\ref{lagrangian1}) implies
processes that violate $L$, like inverse-decays followed by $N_1$ decays
$\ell_1 \leftrightarrow N_1 \leftrightarrow\bar \ell_1$, off-shell $\Delta
L=2$ scatterings $\ell_1 H\leftrightarrow \bar\ell_1 \bar H$, $\Delta L=1$
scatterings involving the top-quark like $N_1 \ell_1 \leftrightarrow Q_3 \bar
t$ or involving the gauge bosons like $N_1\ell_1 \to A \bar H$ (with
$A=W_i,\,B$).  The temperature range in which $\not\!\! L$ processes can be
important for leptogenesis is around $T\sim M_1$.  This is because if the
$\lambda_1$ couplings were large enough that these processes were already
relevant at $T\gg M_1$ (when the Universe expansion is fast) than they would
come into complete thermal equilibrium at lower temperatures (when the
expansion slows down) thus forbidding the survival of any macroscopic $L$
asymmetry.  On the other hand at $T\ll M_1$ decays, inverse decays and $\Delta
L=1$ scatterings are Boltzmann suppressed, $\Delta L=2$ scatterings are power
suppressed, and therefore $L$ violating processes become quite inefficient as
the temperature drops well below $M_1$.

\begin{figure}
  \psfig{figure=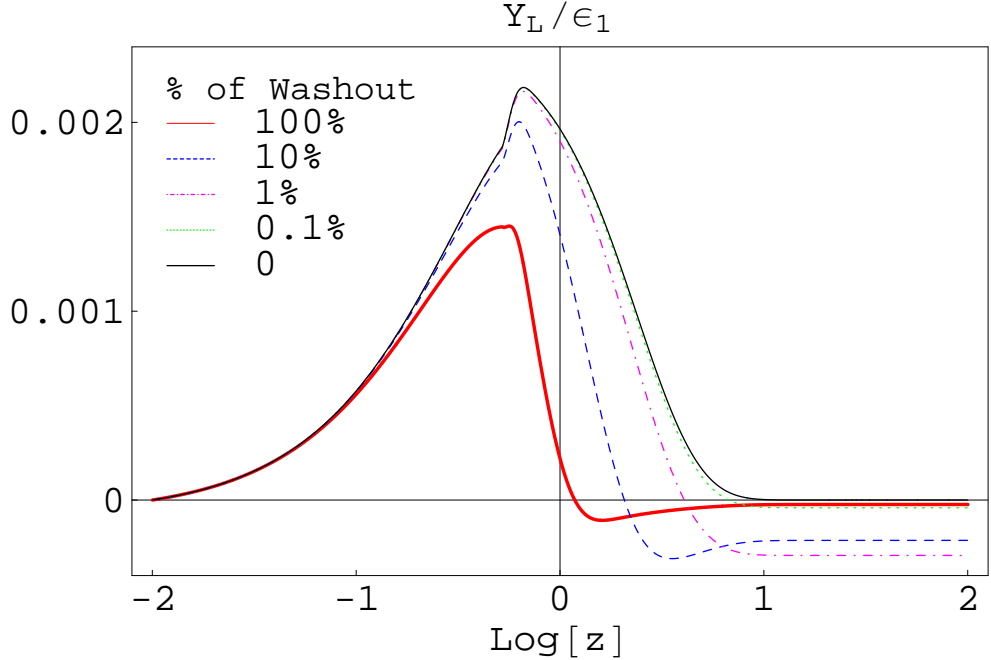,width=15truecm}
\caption{Evolution of the lepton asymmetry plotted against $z=M_1/T$. 
  The different curves depict the effects of reducing progressively the rates
  of the washout processes (as detailed in the legend).  Complete switch off
  of the washouts (thin solid black line) yields a vanishing lepton asymmetry.
}
\label{fig:Ylevolution}
\end{figure}

The possibility of generating an asymmetry between the number of leptons
$n_{\ell_1}$ and antileptons $n_{\bar \ell_1}$ is due to a non-vanishing $CP$
asymmetry in $N_1$ decays:

\begin{equation}
\epsilon_1\equiv \frac{
\Gamma(N_1\to \ell_1 H)-
\Gamma(N_1\to \bar \ell_1 \bar H)}
{\Gamma(N_1\to \ell_1 H)+
\Gamma(N_1\to \bar \ell_1 \bar H)} \neq 0.
\end{equation}
In order that a macroscopic $L$ asymmetry can build up, the condition
that $\not\!\! L$ reactions are (at least slightly) out of equilibrium at
$T\sim M_1$ must also be satisfied. This condition can be expressed in terms of two
dimensionfull parameters, defined in terms of the Higgs vev $v\equiv \langle
H\rangle$ and of the Plank mass $M_P$ as:
\begin{equation}
\label{mstar}
\tilde m_1 =\frac{(\lambda\lambda^\dagger)_{11}\,v^2 }{M_1}, \qquad\qquad \quad  m_* \approx
10^3\,\frac{v^2}{M_P}\approx 10^{-3}\,{\rm eV}.
\end{equation}  
The first parameter ($\tilde m_1$) is related to the rates of $N_1$ processes
(like decays and inverse decays) while the second one ($m_*$) is related to
the expansion rate of the Universe at $T\sim M_1$.  When $\tilde m_1\lsim
m_*$, $\not\!\! L \>$ processes are slower than the Universe expansion rate
and leptogenesis can occur.  As $\tilde m_1$ increases to values larger than
$m_*$, $\not\!\! L$ reactions approach thermal equilibrium thus rendering
leptogenesis inefficient because of the back-reactions that tend to erase any
macroscopic asymmetry.  However, even for $\tilde m_1$ as large as $\sim
100\,m_*$ a lepton asymmetry sufficient to explain the BAU can be generated.
It is customary to refer to the condition $\tilde m_1> m_*$ as to the {\it
  strong washout regime} since washout reactions are rather fast.  This regime
is considered more likely than the {\it weak washout regime} $\tilde m_1< m_*$
in view of the experimental values of the light neutrino mass-squared
differences (that are both $> m_*^2$) and of the theoretical lower bound
$\tilde m_1 \geq m_{\nu_1}$, where $m_{\nu_1}$ is the mass of the lightest
neutrino.  The strong washout regime is also theoretically more appealing
since the final value of the lepton asymmetry is independent of the particular
value of the $N_1$ initial abundance, and also of a possible asymmetry
$Y_{\ell_1}\! =(n_{\ell_1}-n_{\bar \ell_1})/s\neq 0$ (where $s$ is the entropy
density) preexisting the $N_1$ decay era.  This last fact has been often used
to argue that for $\tilde m_1> m_*$ only the dynamics of the lightest Majorana
neutrino $N_1$ is important, since asymmetries generated in the decays of the
heavier $N_{2,3}$ would be efficiently erased by the strong $N_1$-related
washouts.  As we will see below, the effects of $N_1$ interactions on the
$Y_{\ell_{2,3}}$ asymmetries are subtle, and the previous argument is
incorrect.  The result of numerical integration of the Boltzmann equations for
$Y_{\ell_1}$ can be conveniently expressed in terms of an efficiency factor
$\eta_1$, that ranges between 0 and 1:
\begin{equation}
Y_{\ell_1} = 3.9\times 10^{-3}\  \eta_1 \epsilon_1, 
\qquad\qquad \qquad \eta_1 \approx \frac{m_*}{\tilde m_1}.
\end{equation}
The second relation gives a rough approximation for $\eta_1$ in the strong
washout regime, that will become useful in analyzing the impact of flavor
effects.  Clearly, too strong washouts ($\tilde m_1 \gg m_*$) can put in
jeopardy the success of leptogenesis by suppressing too much the efficiency.
However, it should also be stressed that washouts constitute a fundamental
ingredient to generate a lepton asymmetry. This is particularly true in
thermal leptogenesis, with zero initial $N_1$ abundance, and is illustrated in
fig.~\ref{fig:Ylevolution} where the evolution of the lepton asymmetry for a
representative model is plotted against decreasing values of the temperature.
The different curves correspond to different level of (artificial) reduction
in the strength of the washout rates (but not in the $N_1$ production rates)
from the model value (solid red line), to 10\% (dashed blue line), 1\%
(dot-dashed pink line) and 0.1\% (dotted green line).  The solid black line
corresponds to switching off all back-reactions.  (Of course the last four
curves correspond to unphysical conditions.) It is apparent that while a
partial reduction in the washout rates is beneficial to leptogenesis, an
excessive reduction suppresses the final asymmetry and eventually, when
washouts are switched off completely, no asymmetry survives.
This behavior can be understood as follows: all leptogenesis processes can be
seen as scatterings between standard model particle states $X,\,Y$ involving
intermediate on-shell and off-shell unstable $N_1$'s: $X \leftrightarrow
N^{(*)}_1 \leftrightarrow Y$. Since the CP asymmetry of any $X \leftrightarrow
Y$ process is at most of ${\cal O}(\lambda^6)$\cite{Kolb:1979qa}, if the
lepton asymmetries generated in the different processes were exactly
conserved, the overall amount that could survive would not exceed this order.
Moreover, since ${\cal O}(\lambda_1^6)$ asymmetries are systematically
neglected in the Boltzmann equations, the numerical result would be exactly
zero.  However, the on-shell and off-shell components of each process have
much larger CP asymmetries of ${\cal O}(\lambda_1^4)$, and the cancellation to
${\cal O}(\lambda_1^6)$ occurs because they are of opposite sign and (at
leading order in the couplings) of the same magnitude.  Moreover, since the
long range and short range components of each process have different time
scales, at each instant during leptogenesis a lepton asymmetry up to ${\cal
  O}(\lambda_1^4)$ can be present.  Washout processes by definition do not
conserve the lepton asymmetries, and most importantly they act unevenly over
the different processes as well as over their short and long range components,
erasing more efficiently the asymmetries generated in $N_1$ production
processes and off-shell scatterings that on average occur at earlier times,
and washing out less efficiently the asymmetries of processes that destroy
$N_1$'s (on-shell scatterings and decays). It is thanks to the washouts that
an unbalanced lepton asymmetry up to ${\cal O}(\lambda_1^4)$ can eventually
survive. In the next section we will see that when flavor effects are
important, washouts can play an even more dramatic role in leptogenesis.

The possibility of deriving an upper limit for the the light neutrino
masses~\cite{buch02-03} follows from the existence of a theoretical bound on
the maximum value of the $CP$ asymmetry $\epsilon_1$ (that holds when $
N_{1,2,3}$ are sufficiently hierarchical, and $m_{\nu_{1,2,3}}$ quasi
degenerate) and relates $M_1$, $m_{\nu_3}$ and the washout parameter $\tilde
m_1$:
\begin{equation}
\label{limit}
  | \epsilon_1| \leq    \left[\frac{3}{16\pi}\frac{M_1}{v^2} (m_{\nu_3}-m_{\nu_1})\right]
  \,\sqrt{1-\frac{m_{\nu_1}^2}{\tilde m_1^2}}. 
\end{equation}
The term in square brackets is the so called  Davidson-Ibarra limit~\cite{da02}
while the correction in the square root was first given in ref..~\cite{hamby03}
When $m_{\nu_3}\gsim 0.1\,$eV, the light neutrinos are quasi-degenerate and
$m_{\nu_3}-m_{\nu_1}\sim \Delta m^2_{atm}/2 m_{\nu_3} \to 0$ so that, to keep
$\epsilon_1$ finite, $M_1$ is pushed to large values $\gsim 10^{13}\,$GeV.
Since at the same time $\tilde m_1$ must remain larger than $m_{\nu_{1}} $
the washouts also increase, until the surviving asymmetry is
too small to explain the BAU.\footnote{$\Delta L=2$ washout processes, that
  depend on a different parameter than $\tilde m_1$, and that can become
  important when $M_1$ is large, also play a role in establishing the limit.}
The    limit  $m_{\nu_3} \lsim 0.15\,$eV results.

\section{Lepton flavor effects}

In the Lagrangian (\ref{lagrangian1}) the terms involving the charged
lepton Yukawa couplings have not been included.  Since all these
couplings are rather small, if leptogenesis occurs at temperatures $T
\gsim 10^{12}\,$GeV, when the Universe is still very young, not many of
the related (slow) processes could have occurred during its short
lifetime, and leptogenesis has essentially no knowledge of lepton
flavors. At $T\lsim 10^{12}\,$GeV the reactions mediated by the tau
Yukawa coupling  $h_\tau$ become important, and at $ T\lsim 10^{9}\,$GeV
also $h_\mu$-reactions have to be accounted for. 
Including the  Yukawa terms for the leptons  yields the  Lagrangian:
\begin{equation}
\label{lagrangian2}
{\cal L} =\frac{1}{2}\left[\bar N_1 (i\! \not\! \partial) N_1 - 
M_1 N_1 N_1\right] -(\lambda_{1i}\,
\bar N_1\,  \ell_{i}{H} +h_i\bar e_i\ell_i H^\dagger +  {\rm h.c.}),
\end{equation}
where (in the flavor basis) the matrix $h$ of the Yukawa
couplings is diagonal.  The flavor content of the (anti)lepton doublets
$\ell_1$ ($\bar \ell'_1$) to which $N_1$ decays is now important, since these
states do not remain coherent, but are effectively resolved into their flavor
components by the fast Yukawa interactions
$h_i$.~\cite{Barbieri:1999ma,Abada:2006fw,Nardi:2006fx} Note that because of
$CP$ violating loop effects, in general $CP(\bar \ell'_1)\neq \ell_1$, that is
the antileptons produced in $N_1$ decays are not the $CP$ conjugated of the
leptons, implying that the flavor projections $K_i\equiv |\langle\ell_i
|\ell_1 \rangle |^2$ and $\bar K_i\equiv |\langle\bar\ell_i |\bar\ell_1'
\rangle |^2$ differ:  $\Delta K_i = K_i-\bar K_i \neq 0$.  
The flavor $CP$ asymmetries can be defined as:~\cite{Nardi:2006fx}
\begin{equation}
\epsilon_1^i = \frac{
\Gamma(N_1\to \ell_i H)-
\Gamma(N_1\to \bar \ell_i \bar H)}
{\Gamma_{N_1}}=K_i\epsilon_1 + {\Delta K_i}/{2}.
\end{equation}
The factor $\Delta K_i$ in the second equality accounts for the flavor
mismatch between leptons and antileptons. The factor $K_i$ in front of
$\epsilon_1$ accounts for the reduction in the strength of the $N_1$-$\ell_i$
coupling with respect to $N_1$-$\ell_1$, and thus reduces also the strength of
the washouts for the $i$-flavor, yielding an efficiency factor
$\eta_{1}^i=\min (\eta_1/K_i,1)$.  Assuming for illustration $\eta_1/K_i < 1$
the resulting asymmetry is
\begin{equation}
Y_L \approx \sum_i \epsilon^i_1\, \eta_{1}^i \approx n_f 
Y_{\ell_1} + \sum_i\frac{\Delta K_i}{2K_i}\>\frac{m_*}{\tilde m_1}.
\end{equation}
In the first term on the r.h.s. $n_f$ represents the number of flavors
effectively resolved by the charged lepton Yukawa interactions ($n_f=2$ or 3)
while $Y_{\ell_1}$ is the asymmetry that would have been obtained by
neglecting the decoherence of $\ell_1$. The second term, that is controlled by
the `flavor mismatch' factor $\Delta K_i$, can become particularly large in
the cases when the flavor $i$ is almost decoupled from $N_1$ ($K_i \ll 1$).
This situation is depicted in fig.~2 for the two-flavor case and for two
different temperature regimes.  The two flat curves give $|Y_{B-L}|$ as a
function of the flavor projector $K_\tau$ assuming $\Delta K_\tau=0$, and show
rather clearly the enhancement of a factor $\approx 2$ with respect to the one
flavor case (the points at $K_\tau=0,\,1$).  The other two curves are peaked
at values close to the boundaries, when $\ell_\tau$ or a combination
orthogonal to $\ell_\tau$ are almost decoupled from $N_1$, and show how the
$\ell_1$-$\ell'_1$ flavor mismatch can produce much larger enhancements.

It was first noted in ref.~\cite{Nardi:2006fx} that flavored-leptogenesis can
be viable even when the branching ratios for decays into leptons and
antileptons are equal, that is in the limit when $L$ is conserved in decays
and the total asymmetry $\epsilon_1=0$ vanishes.  This is a surprising
possibility, that can occur when the $CP$ asymmetries for the single flavors
are non-vanishing, thanks to the fact that lepton number is in any case
violated by the washout interactions.

In conclusion, the relevance of flavor effects is at least twofold:
\begin{figure}
\hspace{.8cm}
\psfig{figure=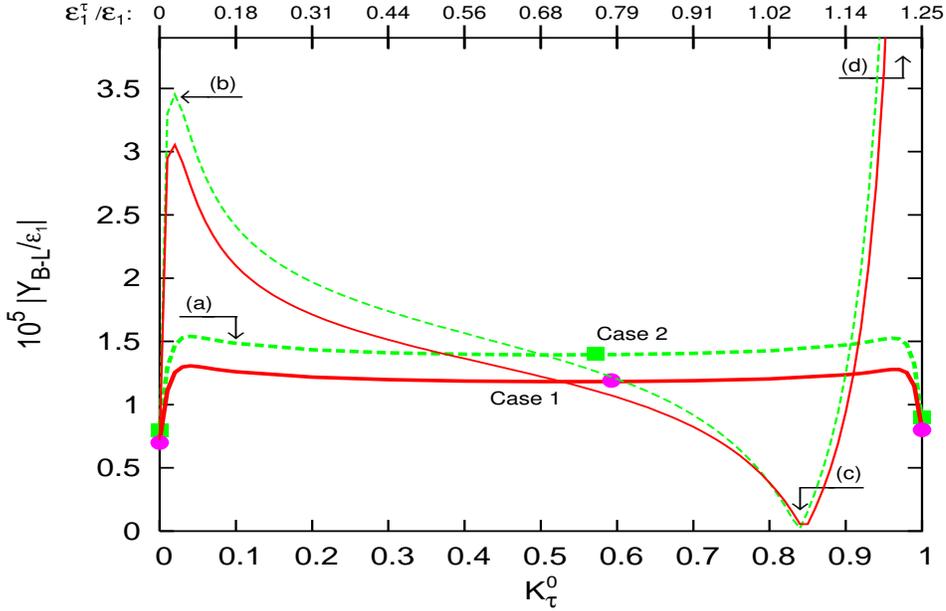,width=.35\textheight,height=13truecm,angle=270}
\caption{
  $|Y_{B-L}|$ (in units of $10^{-5}|\epsilon_1|$) as a function of $K_\tau $
  in two two-flavor regimes.  The thick solid and dashed lines correspond to
  the special case when $K_\tau =\bar K_\tau$. The two thin lines give an example
  of the results for $K_\tau \neq \bar K_\tau$. The filled circles and squares
  at $K_\tau = 0\,,1$ correspond to the aligned cases where flavor effects
are irrelevant.
\label{fig:kplot}}
\end{figure}
\begin{itemize} 
\item[1.] 
The  BAU resulting from leptogenesis can be several times larger 
than what would be obtained neglecting flavor effects. 
\item[2.] 
If leptogenesis occurs at temperatures when flavor
effects are important, the limit on the light neutrino 
masses does not hold.~\cite{Abada:2006fw,DeSimone:2006dd}   
This is because there is no analogous of the Davidson-Ibarra bound  
in eq.~(\ref{limit}) for the flavor asymmetries $\epsilon_1^i$. 

\end{itemize}

\section{The effects of the  heavier Majorana neutrinos}

What about the possible effects of the heavier Majorana neutrinos $N_{2,3}$
that we have so far neglected? A few recent studies analyzed the so called
``$N_1$-decoupling'' scenario, in which the Yukawa couplings of $N_1$ are
simply too weak to washout the asymmetry generated in $N_2$ decays (and
$\epsilon_1$ is too small to explain the BAU).~\cite{decoupling}
This is a consistent scenario in which $N_2$ leptogenesis could successfully
generate enough lepton asymmetry.  However, in the opposite situation when
the Yukawa couplings of $N_1$ are very large, it was generally assumed that
the asymmetries related to $N_{2,3}$ are irrelevant, since they would be
washed out during $N_1$ leptogenesis.  In contrast to this, in
ref.~\cite{Barbieri:1999ma} (and more recently also in
ref.~\cite{Strumia:2006qk}) it was stated that part of the asymmetry from
$N_{2,3}$ decays does in general survive, and must be taken into account when
computing the BAU.  In ref.~\cite{Engelhard:2006yg} Engelhard, Grossman, Nir
and myself carried out a detailed study of the fate of a lepton asymmetry
$Y_P$ preexisting $N_1$ leptogenesis, and we reached conclusions that agree with
these statements. I will briefly describe the reasons for this and the
importance of the results.  Including also $N_{2,3}$ the leptogenesis
Lagrangian reads:
\begin{equation}
\label{lagrangian3}
{\cal L} =\frac{1}{2}\left[\bar N_\alpha (i\! \not\! \partial) N_\alpha - 
{N_\alpha} M_\alpha N_\alpha\right] -(\lambda_{\alpha i}\,
\bar N_\alpha\,  \ell_{i}{H} +h_i\bar e_i\ell_i H^\dagger +  {\rm h.c.}),
\end{equation}
where the heavy neutrinos are written in the mass basis with $\alpha=1,2,3$.
It is convenient to define the three (in general non-orthogonal) combinations
of lepton doublets $\ell_\alpha$ to which the corresponding $N_\alpha$ decay:
\begin{equation}
|{\ell_\alpha}\rangle=(\lambda\lambda^\dagger)_{\alpha\alpha}^{-1/2}
\sum_i\lambda_{\alpha i}\,|{\ell_i}\rangle.
\end{equation}
Let us discuss for definiteness the case when $N_2$-related washouts are not
too strong ($\tilde m_2\not\gg m_*$) , so that a sizeable asymmetry
proportional to $\epsilon_2$ is generated, while $N_1$-related washouts are so
strong that by itself $N_1$ leptogenesis would not be successful ($\tilde m_1\gg m_*$).
%
%
To simplify the arguments, let us also impose two additional conditions:
thermal leptogenesis, that is a vanishing initial $N_1$ abundance
$n_{N_1}(T\gg M_1)\approx0$, and a strong hierarchy $M_2/M_1\gg1$. From this
it follows that there are no $N_1$ related washout effects during $N_2$
leptogenesis and, because $n_{N_2}(T\approx M_1)$ is Boltzmann suppressed,
there are no $N_2$ related washouts during $N_1$ leptogenesis. Thus $N_2$ and
$N_1$ dynamics are decoupled. Now, the second condition in (\ref{conditions})
implies that already at $T\gsim M_1$ the interactions mediated by the $N_1$
Yukawa couplings are sufficiently fast to quickly destroy the coherence of
$\ell_2$ produced in $N_2$ decays.  Then a statistical mixture of $\ell_1$ and
of states orthogonal to $\ell_1$ builds up, and it can be described by a
suitable {\it diagonal} density matrix.  For simplicity, let us assume that
both $N_{2}$ and $N_1$ decay at $T\gsim10^{12}\,$GeV when flavor effects are
irrelevant.  In this regime a convenient choice for the orthogonal lepton
basis is $(\ell_1,\,\ell_{1_\perp}$) where $\ell_{1_\perp}$ denotes generically
the flavor components orthogonal to $\ell_1$.
Then any asymmetry $Y_P$ preexisting the $N_1$ leptogenesis phase 
(as for example $Y_{\ell_2}$) 
decomposes as:
\begin{equation}
\label{eq:c2}
Y_P= Y_{\ell_{1_\perp}} + 
Y_{\ell_1}\,. 
\end{equation}
The crucial point here is that in general we can expect $Y_{\ell_{1_\perp}}$ to be
of the same order than $Y_P$, and since $\ell_{1_\perp}$ is orthogonal to
$\ell_1$, this component of the asymmetry remains protected against $N_1$
washouts.  Therefore, a finite part of any preexisting asymmetry (and in
particular of $Y_{\ell_2}$ generated in $N_2$ decays) survives through $N_1$
leptogenesis.  A more detailed study~\cite{Engelhard:2006yg} reveals also some
additional features.  For example, in spite of the strong $N_1$-related
washouts $Y_{\ell_1}$ is not driven to zero, rather, only the sum of
$Y_{\ell_1}$ and of the Higgs asymmetry $Y_H$ vanishes, but not the two
separately.  (This can be traced back to the presence of a conserved charge
related to $Y_{\ell_{1_\perp}}$.)

For $10^9\lsim M_1 \lsim 10^{12}\,$GeV the lepton flavor structures are only
partially resolved during $N_1$ leptogenesis, and a similar result is
obtained.  However, when $M_1 \lsim 10^9\,$GeV and the full flavor basis
$(\ell_e,\ell_\mu,\ell_\tau)$ is resolved, there are no directions in flavor
space where an asymmetry can remain protected, and then $Y_P$ can be
completely erased independently of its flavor composition.  In conclusion, the
common assumption that when $N_1$ leptogenesis occurs in the strong washout
regime the final BAU is independent of initial conditions, does not hold in
general, and is justified only in the following cases:~\cite{Engelhard:2006yg}
{\it i)}~Vanishing decay asymmetries and/or efficiency factors for $N_{2,3}$
($\epsilon_{{2}}\eta_{2}\approx 0$ and $\epsilon_{{3}}\eta_{3}\approx 0$);
{\it ii)}~$N_1$-related washouts are still significant at $T\lsim10^9\,$GeV;
{\it iii)}~Reheating occurs at a temperature in between $M_2$ and $M_1$.  In
all other cases the initial conditions for $N_1$ leptogenesis, and in
particular those related to the possible presence of an initial asymmetry from
$N_{2,3}$ decays, cannot be ignored when calculating the BAU, and any
constraint inferred from analyses based only on $N_1$ leptogenesis are not
reliable.

\section*{Acknowledgments}
This contribution is based on work done in collaboration with G. Engelhard, Y.
Grossman, Y. Nir, J.  Racker and E. Roulet. I acknowledge partial support 
from INFN in Italy and from Colciencias in Colombia under contract 
No. 1115-333-18739.

\end{document}